\documentclass[pra,twocolumn,showpacs]{revtex4}
\usepackage{graphicx}
\usepackage{bm}

\begin{document}

\title{Theory of Output Coupling for Trapped Fermionic Atoms}

\author{P.A.S.~Pires Filho}
\author{C.L.~Cesar}
\author{L.~Davidovich}

\affiliation{Instituto de F\'\i sica, Universidade Federal do Rio de
Janeiro,  Caixa Postal 68528,
21941-972 Rio de Janeiro, RJ, Brazil\\}

\date{\today}

\begin{abstract}
We develop a dynamic theory of output coupling, for fermionic atoms initially confined in a magnetic trap.  We consider an exactly soluble one-dimensional model, with a spatially localized delta-type coupling between the atoms in the trap and a continuum of free-particle external modes. The transient dynamics of the atoms, as they leave the trap, is investigated in detail. Two important special cases are considered for the confinement potential: the infinite box and the harmonic oscillator.  We establish that in both cases a bound state of the coupled system appears for any value of the coupling constant, implying that the trap population does not vanish in the infinite-time limit.  For weak coupling, the infinite-time spectral distribution of the outgoing atoms exhibits peaks corresponding to the initially occupied energy levels in the trap; the heights of these peaks increase with the energy. As the coupling gets stronger, the infinite-time spectral distribution is displaced towards "dressed energies" of the fermions in the trap. The corresponding dressed states result from the coupling between the unperturbed fermionic states in the trap, mediated by the coupling between these states and the continuum. In the strong-coupling limit, there is a reinforcement of the lowest-energy dressed mode, which contributes to the spectral distribution of the outgoing beam more strongly than the other modes. This effect is especially pronounced for the one-dimensional box, which indicates that the efficiency of the mode-reinforcement mechanism depends on the steepness of the confinement potential. In this case, a quasi-monochromatic anti-bunched atomic beam is obtained. Results for a bosonic sample are also shown for comparison. 
\end{abstract}


\maketitle
\section{Introduction}\label{intro}

The demonstration of the first atom lasers \cite{ketterle1,Kasevich,Martin,phillips,haensch} has led to questions that are reminiscent of those asked when the first optical lasers were put to work. What is the dynamical behavior and the statistics of  the outgoing beam? How monochromatic it is? Atoms offer an interesting twist to these questions, since they may have bosonic or fermionic behavior, while for photons only the bosonic character manifests itself. One may then ask how the statistical properties of the trapped atoms affects the outgoing beam. 

At  zero temperature, one may guess that the behavior of the outgoing beam should be markedly different in the two cases, since for fermions there would be a multitude of populated trapping levels, due to the Pauli exclusion principle, while for a bosonic gas all the atoms would be in the ground state. One expects therefore that fermionic systems should exhibit a richer dynamics, at zero temperature, as compared to bosonic systems, which have been described by one-level models \cite{hope1,savage,moy,jack,breuer,hope2,jeffers} or mean field theories~\cite{walls,shenzle}. Also, coherence properties of fermionic beams are expected to be quite different from their bosonic-beam counterparts. Indeed, while bosonic beams coming from thermal sources exhibit a bunching effect, anti-bunching has already been experimentally demonstrated for electron beams \cite{henny,oliver,kiesel}.

A simple model for a beam of fermionic atoms extracted from a trap was analyzed some years ago \cite{petrosyan}. More recently, the so called input-output formalism developed for photons \cite{collett,gardiner} and applied to bosonic atoms \cite{hope1} was generalized to fermionic species \cite{meystre}. 

One should note that effects concerning the multi-level structure of the trap should also appear in a non-mean field theory for bosonic atom lasers with a non-zero temperature, problem that has been addressed very little, and only within the Popov approximation \cite{griffin} for trapped systems \cite{band,burnett}.

Theoretical work on degenerate fermionic gases has been greatly stimulated by the first propositions of a superfluid BCS-like state~\cite{houbiers,stoof,baranov}, the  obtainment of the first samples of degenerate fermionic gases~\cite{jin,optical}, and some other recent developments~\cite{inguscio,salomon,johnthomas,pitaevskii,regal}. Theoretical studies have concentrated on the analysis of the BCS state~\cite{holland,pethick,bruun,ruost} and its excitation energies~\cite{minguzzi}, as well as on comparisons between fermionic and bosonic properties~\cite{moore,ketterle}.

In the present work, we develop a dynamic theory of output coupling, for fermionic atoms initially confined in a magnetic trap.  The outgoing atoms are considered as free particles. Our method can be easily generalized however to account for a gravitational field. We consider a one-dimensional model, with a spatially localized delta-type coupling between the atoms in the trap and the continuum of external modes. No external replenishment of the trap is considered, so that this model leads to a decay of the population in the trap and a non-stationary outgoing atomic beam.

For an arbitrary confinement potential, we obtain general time-dependent expressions for the atomic operators corresponding to trapped and free atoms, the trapping-level populations, the spectral distribution and the first- and second-order correlation functions of the atomic outgoing beam. From these general expressions, we particularize the results for two important special cases: the infinite box and the harmonic oscillator. In order to identify more clearly the features associated to the fermionic nature of the trapped atoms, we compare our results with the corresponding properties for a bosonic beam. 

Of special interest is the infinite-time spectral distribution of the non-stationary outgoing beam. It differs markedly from the corresponding distribution previously calculated for single-level bosonic models, which leads to a single approximately Lorentzian peak \cite{jeffers}. For weak coupling, we find a distribution that reflects the structure of the discrete levels of the trap, exhibiting peaks that get higher and narrower as the energy increases. As the coupling gets stronger, the atom's spectral distribution is displaced towards a set of new energies that characterize ``dressed states'' of the fermions in the trap. These dressed states result from the coupling between the unperturbed fermionic states in the trap and the untrapped continuum. In the strong-coupling limit, there is a reinforcement of the lowest-energy dressed mode, which contributes to the spectral distribution of the outgoing beam more strongly than the other modes. This effect is especially pronounced for the infinite box, which indicates that the efficiency of the mode-reinforcement mechanism depends on the steepness of the confinement potential. In this case, a striking effect occurs: the fermionic beam becomes quasi-monochromatic, in spite of the large number of energy levels populated in the initial trapped fermionic system.  As expected, the fermionic second-order correlation functions exhibit the property of anti-correlation.  Under these conditions, we predict therefore a quasi-monochromatic anti-bunched fermionic atomic beam.

The paper is organized in the following way. In the next section we introduce the physical model and write down the basic Hamiltonian. The eigenvalue spectrum for this Hamiltonian is analyzed in Sec.~\ref{bound}, where it is shown, for two forms of the trapping potential (infinite box and harmonic oscillator) that the one-dimensional model always exhibits a bound state, for any value of the coupling constant.  In Sec.~\ref{population}, we analyze the non-Markovian behavior of the number of atoms in the trap. The spectral distribution of the outgoing atoms is analyzed in Sec.~\ref{spectrum}. General expressions for the field operators and the correlation functions of the outgoing atoms are derived in Sec.~\ref{correlation}. The corresponding numerical results are presented in Sec.~\ref{numerics}, where comparisons are made between the bosonic and the fermionic cases. Some detailed calculations are referred to a set of two appendices. 

\section{The Model} \label{model}

The physical model considers a single atomic species in a one-dimensional magnetic trap, with an external electromagnetic field inducing transitions between each trapped level and a continuum of non-trapped states. We ignore the effect of the magnetic field on the non-trapped state, assuming for instance that the trapped level corresponds to an electronic spin component $+1$, while the non-trapped state corresponds to the spin component zero. The total spin of the atom, nuclear plus electronic, is assumed to be a half-integer, so that the atom is a fermion and we neglect the small effects due to the nuclear magnetic moment. We model this system by an effective Hamiltonian, with a bilinear coupling involving the field operators for the trapped and untrapped atoms. The one-particle eigenfunctions of the trapping potential are denoted by $\varphi_n(x)$, the corresponding energy levels being given by $\hbar\omega_n$. The untrapped states are identified by the center-of-mass wavefunction $\psi_\xi(x)$, labeled by a continuous parameter $\xi$, with energy $\hbar\omega_{\xi}$.  Thus, if we consider the untrapped atoms as free particles with mass $M$, we have $\hbar\omega_\xi=(\hbar\xi)^2/2M$, and $\hbar\xi$ is the atomic momentum, while $\psi_{\xi}(x)=\exp(i{\xi}x)/\sqrt{2\pi}$  [normalized so that $\int dx \psi_{\xi}^\ast(x)\psi_{\xi'}(x)=\delta(\xi-\xi')$]. If the atoms were under the action of a gravitational field, for instance, then $\xi$ would be an Airy-function index. The coupling is assumed to be spatially localized and is represented by a delta function. While this seems to be a most unphysical assumption, one can imagine a realization where a very tightly focused pair of Raman laser beams, in a weak trap, would induce the electronic transitions. As long as the beams' waists are much shorter than the de Broglie wavelength of the atoms, which could be the case for the lowest-energy states, the assumption of a delta--function coupling can be a good one. 

The effective Hamiltonian is written as
\begin{equation}
\hat H=\hat H_T+\hat H_{F}+\hat H_{C}\,. \label{Hamiltonian}
\end{equation}

In this expression,
\begin{eqnarray}
 \hat H_T &=&\sum_{n}\hbar \omega_n \hat a_n^{\dagger }\hat a_n\,, \label{Hamiltoniant} \\
\hat H_F&=&\int d\xi \hbar\omega_{\xi}\hat b_{\xi}^{\dagger }\hat b_{\xi}\,,  \label{HamiltonianR} 
\end{eqnarray}
where $\hat H_T$ corresponds to the trapped atoms, $\hat a_n$ is the annihilation operator for an atom in the trapping-potential eigenstate $|\varphi_n\rangle$, $\hat H_F$ describes the untrapped atoms, and $\hat b_\xi$ is the operator that annihilates an untrapped atom with wavefunction $\psi_\xi(x)$.

For fermionic atoms, the above operators obey the anti-commutation relations:
\begin{eqnarray}
\{\hat a_n\textrm{,}\hat a^\dagger_{n'}\}&=&\delta_{n,n'}\,, \label{rel1}\\
\{\hat b_\xi\textrm{,}\hat b^\dagger_{\xi'}\}&=&\delta(\xi-\xi')\,, \\
\{\hat a_n\textrm{,}\hat b^\dagger_{\xi}\}&=&0 \label{rel3}\,.
\end{eqnarray}

The coupling part of the Hamiltonian is given, for a general spatial-dependent coupling, by
\begin{equation}
\hat H_{C}=i\hbar\int dx \lambda(x) \hat{\Psi}^{\dagger}(x) \hat{\Phi}(x) + \textrm{H.c.}\,, \label{HamiltonianOC} 
\end{equation}
where the field operators are given by
\begin{equation}\label{sqexp}
\hat{\Psi}(x)=\int d\xi \psi_\xi(x) \hat b_\xi \,, 
\end{equation}
\begin{equation}\label{sqexp2}
\hat{\Phi}(x)= \sum_{n} \varphi_n(x) \hat a_n\,.
\end{equation}

In terms of the operators $\hat b_\xi$ and $\hat a_n$, the interaction $\hat H_C$ may be written as
\begin{equation}\label{genh}
\hat H_C=i\hbar\sum_n\int d\xi\,g_n(\xi)\hat b^\dagger_\xi \hat a_n+\textrm{H.c}\,,
\end{equation}
where 
\begin{equation}\label{genc}
g_n(\xi)=\int dx\,\lambda(x)\psi_\xi(x)\varphi_n(x)\,.
\end{equation}

In the special case of a delta--function coupling $\lambda(x)=\bar{\lambda}\delta(x)$, Eq.~(\ref{genh}) reduces to
\begin{equation}
\hat H_{C}=i\hbar\bar{\lambda}\sum_n \int d{\xi} \psi^{\ast}_{\xi}(0) \varphi_{n}(0) \hat b^{\dagger}_{\xi} \hat a_{n}  +   \textrm{H.c.}\,,  \label{HamiltonianfinalOC}
\end{equation}
which is the interaction used throughout this paper. Without any loss of generality, the coupling constant $\bar\lambda$ is taken to be real. 

This model may be considered as a multilevel extension of other systems considered before, which have a single level interacting with a continuum \cite{cohen2}. 

For sufficiently strong coupling, bound states have been shown to appear in bilinear Hamiltonians involving the interaction of a single mode of the electromagnetic field with a photon  reservoir  \cite{knight}. The corresponding one-dimensional model, with free massive particles as the reservoir, was shown to exhibit a bound state for any value of the coupling constant \cite{jeffers}. We may thus suspect that the above Hamiltonian also exhibits bound states. This is proven in the next Section, for two special cases of the trapping potential.

\section{Diagonalization of the Hamiltonian}\label{bound}

We generalize in this section the procedure adopted in \cite{jeffers} for bosons in a single trapped level coupled to a reservoir of free massive particles. The existence of a multitude of bound levels in our case does not allow one to reach general conclusions concerning the existence of  bound states for any trapping potential. We consider therefore two specific examples, the infinite-box and the harmonic oscillator, and show that the coupling given in Eq.~(\ref{HamiltonianfinalOC}) leads to the existence of a single bound state, for any value of the coupling constant. 

We take the untrapped atoms as free massive particles (no gravitational field), so that $\psi_k(0)=1/\sqrt{2\pi}$ and the Hamiltonian of the system may be written as 
\begin{eqnarray}
\hat H&=&\sum_n \hbar\omega_n \hat a^\dagger_n \hat a_n + \int^{+\infty}_{-\infty}dk\,\hbar\omega_k \hat b^\dagger_k \hat b_k \nonumber \\
&+&\left[\frac{i\hbar\bar{\lambda}}{\sqrt{2\pi}} \sum_n \varphi_n(0) \hat a_n\int^{+\infty}_{-\infty}dk\,  \hat b^\dagger_k + \textrm{H.c.}\right]\,,
\end{eqnarray}
where in this case $\hbar\omega_k=\hbar^2k^2/2M$.

As in \cite{jeffers}, we introduce the even and odd operators $\hat c_k$ and $\hat d_k$, given by
\begin{eqnarray}
\hat c_k&=&\frac{1}{\sqrt{2}}[\hat b_k+\hat b_{-k}] \,, \label{ck1}\\
\hat d_k&=&\frac{1}{\sqrt{2}}[-\hat b_k+\hat b_{-k}] \,.\label{dk1}
\end{eqnarray}

The operators $\hat d_k$ are not coupled to the trap, so we can consider only the operators $\hat c_k$,  and write
\begin{eqnarray}
\hat H&=&\sum_n \hbar\omega_n \hat a^\dagger_n \hat a_n + \int^{+\infty}_{0}dk\,\hbar\omega_k \hat c^\dagger_k \hat c_k \nonumber \\
&+&\left[ \frac{i\hbar\bar{\lambda}}{\sqrt{\pi}} \sum_n \varphi_n(0)\hat a_n \int^{+\infty}_{0}dk\,  \hat c^\dagger_k  + \textrm{H.c.}\right]\,.
\end{eqnarray} 

In order to diagonalize this Hamiltonian, we apply Fano's procedure~\cite{fano,mahan}, introducing the operators 
\begin{equation}\label{Ak}
\hat A_k=\sum_n \alpha_n(k) \hat a_n + \int^{+\infty}_{0} dk' \gamma(k,k') \hat c_{k'}\,,
\end{equation}
so that
\begin{equation}\label{diag1}
\hat H=\int^{+\infty}_{E_\textrm{min}}dk\, \hbar\Omega(k) \hat A^\dagger_k\hat A_k \,,
\end{equation}
where $E_\textrm{min}$ is the lower bound of $\hat H$.

Since the Hamiltonian given by Eqs.~(\ref{Hamiltonian}), (\ref{Hamiltoniant}), (\ref{HamiltonianR}), and (\ref{HamiltonianfinalOC}) is quadratic in the atomic operators, the same procedure holds for fermionic and bosonic atoms. We start by calculating the commutator $[\hat A_k, \hat H]$, which yields two equivalent expressions, obtained by using either Eq.~(\ref{diag1}) or Eq.~(\ref{Hamiltonian}) for $\hat H$:
\begin{eqnarray}
&&[\hat A_k,\hat H]=\hbar\Omega(k)\hat A_k \nonumber \\
&&=\hbar\Omega(k)\left[\sum_n \alpha_n(k) \hat a_n + \int^{+\infty}_{0} dk' \gamma(k,k') \hat c_{k'}\right]\nonumber\\
&&= \sum_n \hbar\omega_n \alpha_n(k) \hat a_n - \frac{i\hbar\bar{\lambda}}{\sqrt{\pi}}  \sum_n \varphi^\ast_n(0) \alpha_n(k) \nonumber \\
&&\quad\times \int^{+\infty}_{0}dk'\,  \hat c_{k'} + \int^{+\infty}_{0}dk' \hbar\omega_{k'}\gamma(k,k') \hat c_{k'} \nonumber \\
&&\qquad+ \frac{i\hbar\bar{\lambda}}{\sqrt{\pi}} \sum_n \varphi_n(0)\hat a_n \int^{+\infty}_{0}dk' \,\gamma(k,k') \,.
\end{eqnarray}

From this equality we obtain the following equations:
\begin{eqnarray}
[\Omega(k)-\omega_n]\alpha_n(k)=\frac{i\bar{\lambda}}{\sqrt{\pi}}\varphi_n(0)\int^{+\infty}_{0}dk'  \gamma(k,k')\label{alpha}\,,\\
\left[\Omega(k)-\omega_{k'}\right]\gamma(k,k')=-\frac{i\bar{\lambda}}{\sqrt{\pi}}\sum_{n'}\varphi^\ast_{n'}(0)\alpha_{n'}(k) \label{gamma}\,.\ \ \ \
\end{eqnarray}

We consider first the negative-energy solutions of these equations, which correspond to bound states. 

\subsection{Bound states}

For bound states, we may set $\Omega(k)=-\mu^2$, $\mu>0$, so that the bound-state energy is $E_B=-\hbar\mu^2$, and let in this case $\alpha_n(k)\rightarrow\alpha_{\mu,n}$, $\gamma(k,k')\rightarrow\gamma_\mu(k')$, so that  Eq.~(\ref{Ak}) is replaced by
\begin{equation}\label{Amu}
\hat A_\mu=\sum_n\alpha_{\mu,n}\hat a_n+\int_0^\infty\gamma_\mu(k)\hat c_k\,dk
\end{equation}
and Eqs.~(\ref{alpha}) and (\ref{gamma}) become:
\begin{eqnarray}
\alpha_{\mu,n}&=&-\frac{i\bar{\lambda}}{\sqrt{\pi}} \frac{\varphi_n(0)}{\mu^2+\omega_n}\int^{+\infty}_0 dk\, \gamma_\mu(k)\label{alphamu}\,, \\
\gamma_\mu(k)&=&\frac{i\bar{\lambda}}{\sqrt{\pi}} \frac{1}{\mu^2+\omega_k}\sum_{n'}\varphi^\ast_{n'}(0)\alpha_{\mu,n'} \,. \label{gammamu}
\end{eqnarray}

Substituting Eq.~(\ref{gammamu}) into Eq.~(\ref{alphamu}), we obtain
\begin{eqnarray}
\alpha_{\mu,n}&=&\frac{\bar{\lambda}^2}{\pi} \frac{\varphi_n(0)}{\mu^2+\omega_n} \int^{+\infty}_0 \frac{dk}{\mu^2+\omega_k} \nonumber \\ 
&\times& \sum_{n'}\varphi^\ast_{n'}(0)\alpha_{\mu,n'} \,. \label{alphamusum}
\end{eqnarray}

Multiplying the last equation by $\varphi^\ast_n(0)$ and summing over $n$, we have
\begin{eqnarray}
\sum_n \varphi^\ast_n(0)\alpha_{\mu,n}&=&\frac{\bar{\lambda}^2}{\pi} \sum_n\frac{|\varphi_n(0)|^2}{\mu^2+\omega_n} \int^{+\infty}_0 \frac{dk}{\mu^2+\omega_k} \nonumber \\ 
&\times&\sum_{n'}\varphi^\ast_{n'}(0)\alpha_{\mu,n'} \,.
\end{eqnarray}

This equation immediately yields the eigenvalue equation for $\mu$:
\begin{equation}
2\bar{\lambda}^2F(\mu^2) I(\mu^2)=1 \label{eqmuap}\,,
\end{equation}
where
\begin{eqnarray}
F(y)&=&\sum_n\frac{|\varphi_n(0)|^2}{y+\omega_n} \label{eqfmuap}\,,\\
I(y)&=&\frac{1}{2\pi}\int_0^\infty \frac{dk}{y+\omega_k}= \sqrt{\frac{M}{2\hbar y}}\,,\label{ifreeap}
\end{eqnarray}
and we have used in Eq.~(\ref{ifreeap}) that $\omega_k=\hbar k^2/2M$.

Replacing Eq.~(\ref{ifreeap}) into Eq.~(\ref{eqmuap}), we get the eigenvalue equation
\begin{equation}
\bar{\lambda}^2\sqrt{\frac{2M}{\hbar}}F(\mu^2)=\mu \label{eqmu2}\,.
\end{equation}

We can see that this equation has one and only one solution if $F(y)$ is finite when $y=0$, and $F(y)\rightarrow0$ when $y\rightarrow\infty$. This will be shown to be the case for the two special cases considered in this paper. One should note however that the form of this eigenvalue equation is highly dependent on the dimensionality of the problem. This dependence is quite apparent in the expression for the function $I(y)$, where the divergence for $y=0$ disappears if one replaces in Eq.~(\ref{ifreeap}) $dk$ by $d^3k$ (adding up a cutoff to the upper integration limit, so that the integral remains finite). This would imply the replacement of $\mu$ on the right-hand side of Eq.~(\ref{eqmu2}) by a function of $\mu$ that would not go to zero when $\mu\rightarrow0$, and therefore the bound state would appear only for a sufficiently strong coupling. For bosons at zero temperature, the dependence on the dimensionality of the bound state of the corresponding Hamiltonian (with just one bound-level)  was explicitly demonstrated in Ref.~\cite{jeffers}. 

The functions $\alpha_{\mu,n}$ and $\gamma_\mu(k)$ may be obtained in the following way. We impose the condition $\{\hat A_\mu,\hat A^\dagger_\mu\}=1$ (for bosons we would replace the anti-commutator by a commutator, with the same results at the end), obtaining
\begin{equation}
\sum_n |\alpha_{\mu,n}|^2 + \int^{+\infty}_0 dk |\gamma_\mu(k)|^2 =1\,.
\end{equation}

Replacing Eqs.~(\ref{gammamu}) and (\ref{alphamusum}) into this equation, we obtain, except for an irrelevant overall phase factor that can be absorbed into the definition of the states $\varphi_n(0)$:
\begin{equation}\label{sumphi}
\sum_n \varphi^\ast_n(0)\alpha_{\mu,n}=\frac{\mu F(\mu^2)}{\sqrt{F(\mu^2)/2-\mu^2F'(\mu^2)}}\,,
\end{equation}
where $F'(\mu^2)$ is the derivative of $F(y)$, defined by Eq.~(\ref{eqfmuap}), evaluated at $y=\mu^2$:
\begin{equation}
F'(\mu^2)=-\sum_n \frac{|\varphi_n(0)|^2}{(\mu^2+\omega_n)^2}\label{eqgmu}\,.
\end{equation}

Taking Eq.~(\ref{sumphi}) into Eqs.~(\ref{gammamu}) and (\ref{alphamusum}), we get finally
\begin{eqnarray}
\alpha_{\mu,n}&=&\frac{\varphi_n(0)\mu/(\mu^2+\omega_n)}{\sqrt{F(\mu^2)/2-\mu^2F'(\mu^2)}}\,,\label{amu}\\
\gamma_\mu(k)&=&\frac{i\bar\lambda\mu F(\mu^2)/\left[\sqrt{\pi}(\mu^2+\omega_k)\right]}{\sqrt{F(\mu^2)/2-\mu^2F'(\mu^2)}}\,.\label{gammamu2}
\end{eqnarray}

We discuss now the solutions of Eq.~(\ref{eqmu2}) for two important special cases of trapping potential: the infinite box and the harmonic oscillator. We show that in both cases there is one and only one bound state, for any non-vanishing value of the coupling constant. This implies that Eq.~(\ref{diag1}) becomes
\begin{equation}\label{diag2}
\hat H=\int_0^{+\infty}dk\,\hbar\omega_k\hat A^\dagger(k)\hat A(k)-\hbar\mu^2\hat A^\dagger_\mu\hat A_\mu\,,
\end{equation}
where $\omega_k=\hbar k^2/2M$, $-\hbar\mu^2$ is the energy of the bound state, and $\hat A_\mu$ is the corresponding  annihilation operator, given by Eq.~(\ref{Amu}).

\subsubsection{Infinite box}

In this case, we have for the trapped particles,
\begin{equation}\label{ebox}
\omega_n=\frac{\hbar\pi^2}{2ML^2}n^2=\omega_1n^2\,,
\end{equation}
the corresponding eigenstates being given, for odd $n$, by:
\begin{equation}\label{statebox}
\varphi_n(r)=\sqrt{\frac{2}{L}}\cos{(k_nr)}\,,
\end{equation}
where $L$ is the length of the box.

For even $n$, the cosine function is replaced by the sine function, which vanishes for $x=0$. The interaction in Eq.~(\ref{HamiltonianfinalOC}) does not couple these states  to the outgoing beam, and they do not contribute to the sum defining $F(\mu^2)$ (this is a consequence of the localized nature of the symmetric coupling). Thus, only odd $n$'s  (even wavefunctions) contribute to $F(\mu^2)$ , which may be written as:
\begin{equation}\label{fmu}
F^{\textrm{(box)}}(\mu^2)=\frac{2}{\omega_1L}\sum_{n \,{\textrm{(odd)}}}\frac{1}{n^2+\mu^2/\omega_1}\,.
\end{equation}

From Eq.~(\ref{finalsum2}) of Appendix \ref{ap1}, we have
\begin{equation}
F^{\textrm{(box)}}(\mu^2)=\frac{\pi}{2L\sqrt{\omega_1}}\frac{\tanh{(\pi\mu/2\sqrt{\omega_1})}}{\mu}\label{resfmubox}\,. 
\end{equation}

Therefore, in this case $F(\mu^2)$ goes to a finite value when $\mu\rightarrow0$, and vanishes when $\mu\rightarrow\infty$. It is clear then that Eq.~(\ref{eqmu2}) has a unique solution. In terms of the adimensional coupling constant $\delta$ defined by 
\begin{equation}\label{delta}
\delta=\bar\lambda\pi^2/L\omega_1\,,
\end{equation}
the weak-coupling limit corresponds to $\delta\ll1$, so that $\mu^2\ll\omega_1$, and  the hyperbolic tangent may be approximated by its value close to the origin, thus yielding
\begin{equation}
E_B=-\hbar\mu^2=-\frac{\delta^4}{(4\pi)^2}\hbar\omega_1\label{muboxweak} \,.
\end{equation}

In the strong-coupling limit $\delta\gg1$, we get $\mu^2\gg\omega_1$, so that the hyperbolic tangent may be approximated  by one, and
\begin{equation}
E_B=-\hbar\mu^2=-\frac{\delta^2}{2\pi^2}\hbar\omega_1\label{muboxstrong}\,.
\end{equation}

\subsubsection{Harmonic oscillator}

For the harmonic trap, we have
\begin{equation}\label{eho}
\hbar\omega_n = \hbar\omega_0 \big(n+1/2\big)
\end{equation}
 and
\begin{eqnarray}
\varphi_n(r) = \big(\frac{1}{\pi d^2} \big)^{1/4} \frac{1}{\sqrt{2^n n!}} e^{-r^2/(2 d^2)} H_n(r/d)\,, \label{wavefuncap}
\end{eqnarray}
where $H_n(x)$ is the Hermite polynomial of order n and $d=\sqrt{\hbar/m\omega_0}$ is the width of the ground state. 

Wave functions corresponding to odd values of $n$ do not contribute to the sum defining $F(\mu^2)$, which becomes now:
\begin{eqnarray}
F^{\hbox{\rm (ho)}}(\mu^2)&=&\frac{1}{2 \omega_0} \sum_{m=0}^{\infty} \frac{|\varphi_{2m}(0)|^2}{m+1/4 +(\mu^2/2\omega_0) } \nonumber \\
&=&\frac{1}{2\omega_0 d}\times\frac{\Gamma(1/4+\mu^2/2\omega_0)}{\Gamma(3/4+\mu^2/2\omega_0)}\,, \label{resfmuho}
\end{eqnarray}
where $\Gamma(x)$ is the Gamma function \cite{abramowitz}. This result is proven in Appendix \ref{ap3}. 

Replacing Eq.~(\ref{resfmuho}) into Eq.~(\ref{eqmu2}), we obtain the final expression for the eigenvalue equation
\begin{equation}\label{eigenh}
2\sqrt{2\omega_0}\delta'^2\frac{\Gamma(1/4+\mu^2/2\omega_0)}{\Gamma(3/4+\mu^2/2\omega_0)}=\mu \label{muho} \,,
\end{equation}
where $\delta'$ is defined by
\begin{equation}\label{delta'}
\delta'=\bar\lambda/2\omega_0 d\,.
\end{equation}
It is easy to verify that Eq.~(\ref{eigenh}) has one and only one solution $\mu>0$ for any $\delta'$. 

In the weak-coupling limit $\delta'\ll 1$, we may neglect the contribution of $\mu$ in the argument of the Gamma functions, thus getting
\begin{equation}
E_B=-\hbar\mu^2=-8\delta'^4\left[\frac{\Gamma(1/4)}{\Gamma(3/4)}\right]^2\hbar\omega_0 \,.\label{muhoweak}
\end{equation}

For strong coupling, we use the following identity~\cite{abramowitz}:
\begin{equation}
z^{b-a}\frac{\Gamma(a+z)}{\Gamma(b+z)}\sim 1+O(z^{-1})\,.
\end{equation}

Identifying $z\rightarrow\mu^2/2\omega_0$, $a\rightarrow1/4$ and $b\rightarrow3/4$, we obtain for $\delta'\gg 1$:
\begin{equation}
E_B=-\hbar\mu^2=-8\delta'{}^2\hbar\omega_0 \,. \label{muhostrong}
\end{equation}

\subsection{Positive-Energy Solutions}

From Eqs.~(\ref{alpha}) and (\ref{gamma}), we can write for $\Omega(k)=\omega_k\ge 0$:
\begin{eqnarray}
\alpha_n(k)&=&\frac{i\bar{\lambda}}{\sqrt{\pi}}\frac{\varphi_n(0)}{\omega_k-\omega_n}\int^{+\infty}_{0}dk'  \gamma(k,k')\label{alphak}\,,\\
\gamma(k,k')&=&-\frac{i\bar{\lambda}}{\sqrt{\pi}}\sum_{n'}\varphi^\ast_{n'}(0)\alpha_{n'}(k)\nonumber\\
&\times&\left[\frac{\textrm{P}}{\omega_k-\omega_{k'}}+Z(k)\delta(\omega_k-\omega_{k'})\right] \label{gammak}\,, 
\end{eqnarray}
where $P$ stands for the principal part, and we have assumed for the moment, in getting Eq.~(\ref{alphak}), that $\omega_k\neq\omega_n$, for any $n$. In these equations $Z(k)$ is a function to be determined. Inserting Eq.~(\ref{alphak}) into Eq.~(\ref{gammak}), we obtain the expression for $Z(k)$:
\begin{equation}
Z(k)=-\frac{\hbar k}{M}\frac{\pi}{\bar{\lambda}^2F(-\omega_k)}\,,\label{zk}
\end{equation}
where $F(y)$ was defined in Eq.~(\ref{eqfmuap}), and we have used that $\delta(\omega_k-\omega_{k'})=(M/\hbar |k|)\delta(k-k')$.

Using Eqs.~(\ref{Ak}), (\ref{alphak}), (\ref{gammak}), and (\ref{zk}), and imposing the condition
\begin{equation}
\{\hat A(k),\hat A^\dagger(k^\prime)\}=\delta(k-k^\prime)\,,
\end{equation}
we obtain  
\begin{equation}
\sum_n \varphi^*_n(0)\alpha_n(k)= \frac{\hbar k}{M}\frac{\sqrt{\pi}}{\bar{\lambda}\sqrt{\pi^2+Z^2(k)}}\,.\label{sumalphak}
\end{equation}
Therefore,
\begin{equation}\label{gamk}
\gamma(k,k^\prime)=\frac{-i(\hbar k/M)}{\sqrt{\pi^2+Z^2(k)}}\left[\frac{\textrm{P}}{\omega_k-\omega_{k'}}+Z(k)\delta(\omega_k-\omega_{k'})\right]
\end{equation}
and
\begin{equation}\label{alnk}
\alpha_n(k)=\frac{\bar{\lambda}}{\sqrt{\pi}}\frac{\varphi_n(0)}{\omega_k-\omega_n}\frac{Z(k)}{\sqrt{\pi^2+Z^2(k)}}\,.
\end{equation}

From Eqs.~(\ref{eqfmuap}) and (\ref{zk}), it is easy to check that $\alpha_n(k)$, given by Eq.~(\ref{alnk}), remains finite when $\omega_k\rightarrow\omega_n$. This allows one to remove the restriction $\omega_k\neq\omega_n$, used to get Eq.~(\ref{alphak}), and adopt Eq.~(\ref{alnk}) as the expression for $\alpha_n(k)$ for all values of $k$.

\section{Population in the trap: non-Markovian behavior}\label{population}

One of the consequences of the existence of the bound state is the failure of the Born-Markov approximation for this problem. A related consequence is that a fraction of the atoms remains in the cavity, even in the infinite-time limit $t\rightarrow\infty$. This can be seen by writing down the decomposition of each cavity mode in terms of the eigenmodes of the Hamiltonian:
\begin{equation}\label{at}
\hat a_n(t)= \int_0^\infty \alpha_n^*(k)\hat A_k e^{-i\hbar k^2t/2M}dk+\alpha^*_{\mu,n}e^{i\mu^2t}\hat A_\mu\,,
\end{equation}
and replacing the operators $\hat A_k$ and $\hat A_\mu$ by their expressions in terms of the operators $\hat a_n(0)$ and $\hat c_k(0)$. One gets then:
\begin{eqnarray}\label{at2}
\hat a_n(t)&=&\sum_{n'}\bigg[\int_0^\infty \alpha_n^*(k)\alpha_{n'}(k)e^{-i\hbar k^2t/2M}dk\nonumber\\
&+&\alpha^*_{\mu,n}\alpha_{\mu,n'}e^{i\mu^2t}\bigg]\hat a_{n'}(0)\nonumber\\
&+&\int_0^\infty dk'\bigg[\int_0^\infty dk\,\alpha_n^*(k)\gamma(k,k')e^{-i\hbar k^2t/2M}\nonumber\\
&+&\alpha^*_{\mu,n}\gamma_\mu(k')e^{i\mu^2t}\bigg]\hat c_{k'}(0)\,.
\end{eqnarray}

This expression exhibits explicitly the coupling between the trap modes, which is induced by the coupling with the external modes. If initially only the trap modes are populated, and if one is interested only in normal-ordered correlation functions, the contribution of the operators $\hat c_k(0)$ may be ignored. This will be always the case in the present paper. 

If at time $t=0$ only the cavity mode $n$ is populated, the fraction of atoms left in the same mode at a later time $t$ is given by
\begin{eqnarray}\label{pop}
&{}&\frac{\langle \hat a_n^\dagger(t)\hat a_n(t)\rangle}{\langle\hat a_n^\dagger(0)\hat a_n(0)\rangle}\nonumber\\
&{}&=\left|\int_0^\infty|\alpha_n(k)|^2e^{-i\hbar k^2 t/2M}dk+|\alpha_{\mu,n}|^2e^{i\mu^2t}\right|^2.
\end{eqnarray}

The integral vanishes in the infinite-time limit, since $\alpha_n(k)$ remains finite for all values of $k$, and therefore in this limit the fraction of atoms left in the cavity is $|\alpha_{\mu,n}|^4$. This result is easy to understand: in order to get the residual population, one must multiply the fraction of atoms in the initial mode that are in the bound mode $\hat A_\mu$, given by $|\alpha_{\mu,n}|^2$, by the fraction of the cavity mode $\hat a_n$ present in $\hat A_\mu$, which is also given by $|\alpha_{\mu,n}|^2$. Furthermore, the time-dependent population exhibits oscillations, resulting from the beating between the integral and the discrete contribution in Eq.~(\ref{pop}).

A similar behavior holds if initially more than one bound mode is populated, as it is the case for trapped fermions at zero temperature. The residual population of level $n$ is then given by
\begin{equation}\label{persistent}
\langle\hat a^\dagger_n\hat a_n\rangle(\infty)=|\alpha_{\mu,n}|^2\sum_{n'}|\alpha_{\mu,n'}|^2\langle\hat a_{n'}^\dagger(0)\hat a_{n'}(0)\rangle\,,
\end{equation}
where now $\sum_{n'}|\alpha_{\mu,n'}|^2\langle\hat a_{n'}^\dagger(0)\hat a_{n'}(0)\rangle$ is the fraction of the initial population that is in the bound mode $\hat A_\mu$, and $|\alpha_{\mu,n}|^2$ is the fraction of the cavity mode $\hat a_n$ present in the bound mode.

An expression for the total residual population $N(\infty)$ inside the trap may be obtained from Eqs.~(\ref{persistent}) and (\ref{amu}):
\begin{eqnarray}\label{totalpop}
N(\infty)&=&\sum_n\langle\hat a^\dagger_n\hat a_n\rangle(\infty)=\sum_{n'}|\alpha_{\mu,n'}|^2\langle\hat a_{n'}^\dagger(0)\hat a_{n'}(0)\rangle\nonumber\\
&\times&\frac{2\mu^2F'(\mu^2)}{2\mu^2F'(\mu^2)-F(\mu^2)}\,, 
\end{eqnarray}
where $F'(\mu^2)$ is, as before, the derivative of $F(y)$, given by Eq.~(\ref{eqfmuap}), evaluated at $y=\mu^2$. 

For $N$ bosons at zero temperature, only the term with $n'=0$ contributes to the above sum, and $\langle\hat a_{n'}^\dagger(0)\hat a_{n'}(0)\rangle=N$. On the other hand, for fermions at zero temperature, the initial  population is the same for all levels (one atom for each level, since all the trapped atoms have the same spin), up to the last occupied one (Fermi surface). If the number of atoms is much larger than one, then one may approximate the sum in the above expression by one with an infinite number of terms. The resulting number is an upper bound for the residual population inside the trap, which is actually achieved when $N\rightarrow\infty$:
\begin{equation}\label{totalpopmax}
N_{\textrm{max}}(\infty)=\left[\frac{2\mu^2F'(\mu^2)}{2\mu^2F'(\mu^2)-F(\mu^2)}\right]^2\,.
\end{equation}

In the weak-coupling limit ($\delta\ll1$), an approximate expression for (\ref{totalpop}) may be obtained,  for the 1-D box and the 1-D harmonic oscillator, by using the  results obtained before for the function $F(\mu^2)$ and for the bound-state energy $E_B=-\hbar\mu^2$. We get thus, for the 1-D box,
\begin{equation}
N(\infty)=\frac{\delta^8}{96}\sum_{n\textrm{(odd)}}\frac{\langle\hat a_{n}^\dagger(0)\hat a_{n}(0)\rangle}{n^4}\,,
\end{equation}
and for the harmonic oscillator
\begin{eqnarray}
&&N(\infty)=64\sqrt{\pi}\delta'^8\left[\frac{\Gamma(1/4)}{\Gamma(3/4)}\right]^3\nonumber\\
&&\quad\times\sum_{n=0}^{\infty}\frac{(2n)!}{2^{2n}(n!)^2(n+1/4)^2}\langle\hat a_{2n}^\dagger(0)\hat a_{2n}(0)\rangle\,.
\end{eqnarray}

From these results, we can see that the residual population inside the trap is very small in the weak-coupling limit, being proportional, both for the box and the harmonic oscillator, to the eighth power of the corresponding dimensionless coupling constant.

In the strong-coupling limit ($\delta\gg1$), we get both for the box and the harmonic oscillator that the upper bound for the population inside the trap is $N_{\textrm{max}}(\infty)=1/4$. This result, which is actually achieved when the number of atoms is  much larger than one, shows that a substantial fraction of the atoms remains in the trap in the infinite-time limit. This is a direct consequence of the existence of a bound state of the total Hamiltonian. 

We proceed now to  the calculation of the spectral distribution of the outgoing atomic beam.

\section{Spectral distribution of the outgoing beam}\label{spectrum}

A time-dependent spectral distribution for the outgoing fermionic beam can be obtained from the expression of the free-atom operators in terms of the operators that diagonalize the Hamiltonian:
\begin{equation}\label{ck2}
\hat c_k(t)=\int_0^\infty dk'\gamma^*(k',k)e^{-i\omega_{k'}t}\hat A_{k'}(0)+\gamma^*_\mu(k)e^{i\mu^2t}\hat A_\mu(0)\,.
\end{equation}

From Eqs.~(\ref{Ak}), (\ref{Amu}), and (\ref{ck2}), we get, ignoring the contribution of the operators $\hat c_k(0)$ (since initially the outside modes are empty, and only normal-ordered correlation functions are considered):
\begin{eqnarray}\label{ckt}
\hat c_k(t)&=&\sum_n\bigg[\int_0^\infty dk'\gamma^*(k',k)\alpha_n(k')e^{-i\omega_{k'}t}\nonumber\\
&+&\gamma_{\mu}^*(k)\alpha_{\mu,n}e^{i\mu^2t}\bigg]\hat a_n(0)\,.
\end{eqnarray}

The time-dependent spectral distribution is given by $\langle\hat b_k^\dagger(t) \hat b_k(t)\rangle$. This quantity can be expressed in terms of $\hat c_k(t)$, using that from Eqs.~(\ref{ck1}) and  (\ref{dk1}), 
\begin{eqnarray}\label{bk1}
\hat b_k&=&{1\over\sqrt{2}}(\hat c_k-\hat d_k)\,,\nonumber\\
\hat b_{-k}&=&{1\over\sqrt{2}}(\hat c_k+\hat d_k)\,,
\end{eqnarray}
and that $\hat d_k$ does not couple with the trapped-atoms operators, so that it can be ignored when calculating normal-ordered correlation functions [of course, its presence in Eq.~(\ref{bk1}) is important to get the correct commutation relations for the operators $\hat b_k$ and $\hat c_k$]. One gets then that the time-dependent spectral distribution is given by $\langle\hat c_k^\dagger(t) \hat c_k(t)\rangle/2$.

The integral in Eq.~(\ref{ckt}) can be calculated by using Eqs.~(\ref{gamk}) and (\ref{alnk}). For finite times, one has to consider the contributions from the complex poles of the integrand, which give rise to exponentially decaying terms. These contributions can be handled numerically. An example will be given in Section \ref{numerics}. 

An analytic expression can be obtained in the infinite-time limit. Since the contributions from the complex poles of the integrand in Eq.~(\ref{ckt}) give rise to exponentially decaying terms, they will be negligible in this limit, so the relevant contributions come from the principal part and the delta function in Eq.~(\ref{gamk}).  One gets then:
\begin{eqnarray}\label{cktin}
\hat c_k(t)&\rightarrow&\sum_n\bigg[\frac{-\bar\lambda\varphi_n(0)/\sqrt{\pi}}{\omega_k-\omega_n}\frac{e^{-i\omega_kt}}{\bar\lambda^2(M/\hbar k)F(-\omega_k)+i}\nonumber\\
&+&\gamma_{\mu}^*(k)\alpha_{\mu,n}e^{i\mu^2t}\bigg]\hat a_n(0)\,.
\end{eqnarray}

We will show in the following that the first term on the right-hand side of this equation leads, in the weak- and strong- coupling limits, to narrow peaks, centered around the unperturbed energies of the trapped atoms in the weak-coupling case, and around dressed energies of the coupled atoms in the strong-coupling limit. The last term on the right-hand side of this equation is  the bound-state contribution. It can be shown to be much smaller than the remaining terms in Eq.~(\ref{cktin}) in the regions of  the spectral distribution close to the peaks, so it will be neglected from now on. One gets then, using Eq.~(\ref{bk1}) and neglecting the contribution from $\hat d_k$:
\begin{eqnarray}\label{bkas}
\hat b_k&\approx& \sum_n\left[\frac{-\bar\lambda\varphi_n(0)/\sqrt{2\pi}}{\omega_k-\omega_n}\frac{e^{-i\omega_kt}}{\bar\lambda^2(M/\hbar |k|)F(-\omega_k)+i}\right]\nonumber\\
&\times&\hat a_n(0)\,.
\end{eqnarray}
In view of Eq.~(\ref{bk1}), this equation yields $\hat b_k$ for both signs of $k$.

From this expression, and assuming that the initial state is diagonal in the number representation, one gets the outgoing beam spectral distribution in the infinite-time limit:
\begin{eqnarray}\label{gspectrum}
\langle\hat b^\dagger_k\hat b_k\rangle(\infty)&=&\sum_n\frac{\bar\lambda^2|\varphi_n(0)|^2/\left[2\pi(\omega_k-\omega_n)^2\right]}{\left[1+\bar\lambda^4(M/\hbar k)^2F^2(-\omega_k)\right]}\nonumber\\
&\times&\langle\hat a^\dagger_n(0)\hat a_n(0)\rangle\,.
\end{eqnarray}

From Eqs.~(\ref{zk}) and (\ref{alnk}), it is easy to see that
\begin{equation}
\langle\hat b^\dagger_k\hat b_k\rangle(\infty)={1\over2}\sum_n |\alpha_n(k)|^2\langle\hat a^\dagger_n(0)\hat a_n(0)\rangle\,,
\end{equation}
which shows that the contribution to the spectral distribution from trap level $n$ is proportional to the probability $|\alpha_n(k)|^2$ that the atom in this level is in the eigenmode of the total Hamiltonian with energy $\omega_k$.

We consider now the specialization of Eq.~(\ref{gspectrum}) to the infinite-box and harmonic oscillator potentials. 

\subsection{Infinite Box}\label{sbox}

From Eq.~({\ref{finalsum}), we get:
\begin{equation}\label{fomega}
F^{\hbox{\rm (box)}}(-\omega_k)={\pi^2\over2\omega_1L}{\tan(kL/2)\over kL}\,.
\end{equation}

Taking this result, plus Eqs.~(\ref{ebox}), (\ref{statebox}), and (\ref{delta}) into Eq.~(\ref{bkas}), we get, for large times:
\begin{equation}\label{bas2}
\hat b_k(t)\approx G(k,t)\sum_{n=1 \hbox{\rm (odd)}}^{+\infty} \frac{\delta\sqrt{L/\pi}(kL)^2}{(kL)^2-\pi^2n^2}  \hat a_n(0)\,,
\end{equation}
where
\begin{equation}\label{gkl}
G(k,t)=\frac{-\cos{(|kL/2|)}e^{-i\hbar k^2t/2M}}{i(kL)^2\cos{(|kL/2|)}+(\delta^2/4)\sin{(|kL/2|)}}\,.
\end{equation}
One should note that the singularities in the sum are canceled out by the numerator of $G(k,t)$. 

We discuss now the behavior of these expressions in two limiting cases, corresponding to weak ($\delta\ll1$) and strong ($\delta\gg1$) coupling.
For weak coupling ($\delta\ll1$), the term proportional to $\cos(|kL/2|)$ dominates in the denominator of $G(k,t)$, which exhibits sharp resonances close to values of $k$ that correspond to the bound states of the infinite-box potential: $kL=n\pi$, $n$ odd -- these are the zeros of $\cos(|kL/2|)$. We may thus approximate the expression in Eq.~(\ref{bas2}) by setting, around each peak, $kL=n\pi+\beta_n$, $|\beta_n|\ll1$, keeping only the lowest-order terms in the expansions of the trigonometric functions in Eq.~(\ref{gkl}), and neglecting small corrections in $\beta_n$ for the other $k$-dependent contributions. Neglecting these corrections means that deviations from the Lorentzian shape will be ignored here. 

One gets then, from Eqs.~(\ref{bas2}) and (\ref{gkl}), that, asymptotically in time,
\begin{equation}\label{bas3}
\hat b_k(t)\approx\sum_n \hat b_k^{(n)}(t)\,,
\end{equation}
where 
\begin{equation}\label{baprox}
\hat b_k^{(n)}(t)\approx\frac{i\delta\sqrt{L/\pi}e^{-i\hbar k^2t/2M}/2\pi n}{(kL-n\pi)+2i(\delta/2n\pi)^2}\hat a_n(0)\,.
\end{equation}

The infinite-time spectral distribution is then given by the sum of the contributions from all peaks:
\begin{eqnarray}\label{weaksp}
\langle \hat b_k^\dagger \hat b_k\rangle(\infty)/L&=&\sum_{n\textrm{(odd)}}\frac{\delta^2}{4\pi^3n^2} \frac{1}{(kL-n\pi)^2+(\delta/\sqrt{2}n\pi)^4}\nonumber\\
&\times& \langle \hat a^\dagger_n(0) \hat a_n(0)\rangle\,.
\end{eqnarray}

Under the above approximations, each term in the sum is a Lorentzian with  width $(\Delta kL)_{\textrm{FWHM}} = 2(\delta/n\pi)^2$, and height increasing with the square of $n$. Therefore, the peaks in the spectral distribution become higher and narrower as $n$ increases, in such a way that the area under each peak is proportional to the state population $\langle \hat a^\dagger_n(0) \hat a_n(0)\rangle$ corresponding to the same value of $n$. One should note that the contribution for each peak stems, in this case, from a single bound state. The total number of peaks is equal to the number of initially populated states.

For strong coupling, $\delta\gg1$, the term proportional to $\sin(|kl/2|)$ dominates in the denominator of Eq.~(\ref{gkl}), implying that the peaks are shifted towards the values of $k$ corresponding to the zeroes of $\sin(|kL/2|)$. These values, given by $kL=2m\pi$, $m$ integer, are precisely midway between the weak-coupling peaks, and are associated to dressed energies of the system, which originate from the interaction between the discrete states through the continuum, as shown in Eq.~(\ref{at2}). Indeed, it is easy to see that,  in this case,  the sum over $n$ must be kept in the approximation analog to Eq.~(\ref{baprox}) of the expression in Eq.~(\ref{bas2}) around $kL=2m\pi$, $m$ integer. Therefore, several trap states contribute now to each peak in the spectral distribution. 

The complete infinite-time spectral distribution is given, as before, by the sum of  the contributions for all peaks:
\begin{eqnarray}\label{strsp}
&&\langle \hat b_k^\dagger \hat b_k\rangle(\infty)/L=\sum_{m=1}^\infty\frac{1024/\pi\delta^2}{(4\sqrt{2}m\pi/\delta)^4+(kL-2m\pi)^2}\nonumber\\
&&\qquad\qquad\times\sum_{n=1 \textrm{(odd)}}^{+\infty}\frac{m^4}{(4m^2-n^2)^2}\langle \hat a^\dagger_n(0) \hat a_n(0)\rangle\,.
\end{eqnarray}

This expression differs remarkably from the one in Eq.~(\ref{weaksp}). There is now an infinite number of peaks. Each peak, with width $(\Delta kL)_{\textrm{FWHM}} = 32(m\pi/\delta)^2$, is now fed by all the populations in the trap. For the peak at $kL=2m\pi$, the strongest contributions come from the two populations with $n=2m\pm1$ (first neighbors of $kL=2m\pi$), with heights proportional to $1/(4m+1)^2$. The contributions from the second neighbors is proportional to $1/9(4m+3)^2$, which is less than nine times smaller. The contributions from states that are farther away decrease as $1/n^4$. This implies that the height of the peaks decreases as $1/m^2$, while its width increases as $m^2$, and that the contribution for each peak comes mainly from the populations of the first neighboring trap states. These two states are the main components of the dressed state that contributes to this peak. 

\subsection{Harmonic Oscillator}\label{sho}

Setting $\mu^2=-\omega_k$ in Eq.~(\ref{resfmuho}), we get:
\begin{equation}
F^{\hbox{\rm (ho)}}(-\omega_k)={1\over2\omega_od}{\Gamma\left[1/4-(kd/2)^2\right]\over \Gamma\left[3/4-(kd/2)^2\right]}\,.
\end{equation}

Inserting this result, plus Eq.~(\ref{eho}) into Eq.~(\ref{bkas}), we get:
\begin{equation}\label{boas}
\hat b_k(t)\approx\frac{\delta' d}{\sqrt{2\pi}}\sum_{m=0}^{+\infty}\frac{\varphi_{2m}(0)|kd/2|J(kd,t)}{(kd/2)^2-(m+1/4)}\hat a_{2m}(0)\,,
\end{equation}
where 
\begin{equation}
J(y,t)=\frac{\Gamma^{-1}(1/4-y^2/4)^2)e^{-i\hbar k^2t/2M}}{\delta'^2 \Gamma^{-1}(3/4-y^2/4)+i|y/2|\Gamma^{-1}(1/4-y^2/4)}\,.
\end{equation}

The spectral distribution of the outgoing beam is given by $\langle \hat b_k^\dagger \hat b_k\rangle(\infty)$. From Eq.~(\ref{boas}), we can see that, for $\delta'\ll1$, the peaks of the spectral distribution should be close to the zeroes of $\Gamma^{-1}[1/4-(kd/2)^2]$, while for $\delta'\gg1$ they approach the zeroes of $\Gamma^{-1}[3/4-(kd/2)^2]$, which are, as in the infinite-box case, midway between the weak-coupling peaks. These zeroes correspond to the dressed energies of the harmonic trap, under strong-coupling conditions.  

We obtain now approximate expressions for the spectral distribution in the weak- and strong-coupling regions, by expanding the Gamma functions about the values of $k$ that correspond to the spectral peaks. A useful equality for this purpose is 
\begin{equation}
\Gamma^{-1}(x)=\Gamma(1-x)\frac{\sin{(\pi x)}}{\pi}\, .
\end{equation}

For weak coupling, the peaks are around $(kd/2)^2=m+1/4$, $m$ integer. We get then, by expanding the functions of $k$ in Eq.~(\ref{boas}) around these values, using Eq.~(\ref{wavefuncap}), and approximating the spectral distribution by the sum of the contributions from all the peaks:
\begin{eqnarray}\label{weakspoh}
\langle \hat b^\dagger_k \hat b_k\rangle(\infty)/d&=&\frac{\delta'^2}{2\pi^{7/2}}\sum_m\frac{(2m)!/2^{2m}(m!)^2}{[(kd/2)^2-m-1/4]^2+\tilde{\Gamma}_m^2}\nonumber\\
&\times&\langle \hat a^\dagger_{2m}(0)\hat a_{2m}(0)\rangle\,,
\end{eqnarray}
where the linewidth $\tilde{\Gamma}_m$ is given by
\begin{equation}
\tilde{\Gamma}_m=\frac{\delta'^2}{\pi}\frac{\Gamma\big(m+\textrm{$1\over2$}\big)}{m!\sqrt{m+\textrm{$1\over4$}}}\,.
\end{equation}

For $m\gg1$, we find, using Stirling's approximation, that the linewidth of the peak of order $m$ is given by
\begin{equation}
\tilde{\Gamma}_m= \frac{\delta'^2}{\pi m}\,,
\end{equation}
while the corresponding height is $m^{3/2}/2\pi^2\delta'^2$. 
 
In the strong-coupling regime the peaks are around the values $(kd/2)^2=m+3/4$, so that one gets, approximating the contribution around each peak:
\begin{eqnarray}\label{strspoh}
\langle \hat b^\dagger_k \hat b_k\rangle(\infty)/d&=& \frac{1}{2\delta'^2\pi^{7/2}}\sum_{m=0}^\infty\sum_{n=0}^{\infty}\frac{\big(m+\textrm{$3\over4$}\big)\Gamma^2\big(m+\textrm{$3\over2$}\big)}{\big(m-n+\textrm{$1\over2$})^2}\nonumber\\
&\times&\frac{(2n)!/2^{2n}(n!m!)^2}{[(kd/2)^2-m-\textrm{$3\over4$}]^2+\tilde{\Gamma'}_m^2}\nonumber\\
&&\,\,\,\,\,\times\langle \hat a^\dagger_{2n} \hat a_{2n}(0) \rangle\,,
\end{eqnarray}
where the linewidth is now given by
\begin{equation}
\tilde{\Gamma'}_m=\frac{1}{\delta'^2\pi}\frac{\sqrt{m+\textrm{$3\over4$}}\Gamma\big(m+\textrm{$3\over2$}\big)}{m!}\,.
\end{equation}

One should note that, in the strong-coupling limit, all trapped-level populations contribute to each resonance, as opposed to the weak-coupling limit, when each resonance is associated with a single trapping level. The same phenomenon occurred in the infinite-box potential.

In this case, for $m\gg1$, we have for the linewidth of the peak of order $m$,
\begin{equation}
\tilde{\Gamma'}_m= \frac{m}{\delta'^2\pi}\,,
\end{equation}
while the corresponding height is proportional to $\delta'^2/\pi^2\sqrt{m}$.

\section{Field operators and correlation functions}\label{correlation}

\subsection{Field operators}\label{fo}

The time-dependent field operators for the outgoing atoms are given by:
\begin{equation}\label{fop}
\hat\Psi(x,t)=\int_{-\infty}^{+\infty}dk\,{e^{ikx}\over\sqrt{2\pi}}\,\hat b_k(t)\,.
\end{equation}

From Eqs.~(\ref{ckt}) and (\ref{bk1}), this can be written in the following form, if we ignore the vacuum terms proportional to $\hat c_k(0)$ and $\hat d_k(0)$:
\begin{equation}\label{operator}
\hat\Psi(x,t)=\sum_n N(n,x,t)\hat a_n(0)\,,
\end{equation}
where
\begin{eqnarray}\label{Nfunc}
N(n,x,t)&=&\int_{0}^{+\infty}{dk\over\sqrt{\pi}}\bigg[\int_0^\infty dk'\gamma^*(k',k)\alpha_n(k')\nonumber\\
&\times&e^{-i\omega_{k'}t}+\gamma_{\mu}^*(k)\alpha_{\mu,n}e^{i\mu^2t}\bigg]e^{ikx}.
\end{eqnarray}

We may call this the source contribution to the field operators. One should note that $N(n,x,0)=0$, as expected (no contribution, at $t=0$, of the trap modes to the field operators corresponding to the outgoing atoms). Indeed, from Eqs.~(\ref{at}) and (\ref{ck2}), $\{\hat c_k,\hat a_n^\dagger\}=0$ implies that 
\begin{equation}
\int_0^\infty dk'\gamma^*(k',k)\alpha_n(k')+\gamma_{\mu}^*(k)\alpha_{\mu,n}=0\,.
\end{equation}

From Eqs.~(\ref{operator}) and (\ref{Nfunc}), we can see that the field operator corresponding to the outgoing atoms is given by the sum of two contributions, besides the terms proportional to $\hat c_k(0)$ and $\hat d_k(0)$,
\begin{equation}
\hat\Psi(x,t)=\hat\Psi^{\hbox{\rm (bound)}}(x,t)+\hat\Psi^{\hbox{\rm (run)}}(x,t)\,,
\end{equation}
where the bound-state contribution is given by
\begin{eqnarray}
\hat\Psi^{\hbox{\rm (bound)}}(x,t)&=&\sum_n\int_{-\infty}^{+\infty}{dk\over2\sqrt{\pi}}\gamma_{\mu}^*(k)\alpha_{\mu,n}e^{ikx}e^{i\mu^2t}\nonumber\\
&\times&\hat a_n(0)\,,
\end{eqnarray}
and the running-wave part is
\begin{eqnarray}
\hat\Psi^{\hbox{\rm (run)}}(x,t)&=&\sum_n\int_{0}^{+\infty}{dk\over\sqrt{\pi}}\int_0^\infty dk'\gamma^*(k',k)\alpha_n(k')\nonumber\\
&\times&e^{-i\omega_{k'}t}\cos kx\,\hat a_n(0)\,.
\end{eqnarray}

 The bound-state contribution is readily calculated, by using Eqs.~(\ref{eqmu2}), (\ref{amu}) and (\ref{gammamu2}):
\begin{eqnarray}
\hat\Psi^{\hbox{\rm (bound)}}(x,t)&=&\frac{i\sqrt{\pi}(\mu/\bar\lambda)e^{-\sqrt{2M/\hbar}\mu x}e^{i\mu^2t}}{F(\mu^2)-\mu^2F'(\mu^2)}\nonumber\\
&\times&\sum_n\frac{\varphi_n(0)}{\mu^2+\omega_n}\hat a_n(0)\,,
\end{eqnarray}
which exhibits a spatial dependence that decays exponentially, with a decay constant given by $\sqrt{2M|E_{B}|}/\hbar$, where $E_{B}=-\hbar \mu^2$ is the bound-state energy. 

For the running-wave part, simple results can be obtained by replacing directly into Eq.~(\ref{fop}) the asymptotic results obtained for the operators $\hat b_k$ in the weak- and strong-coupling limit, for the special cases of the infinite box and the harmonic oscillator. We restrict ourselves here to the infinite-box case, in the weak-coupling limit, since the results for the strong-coupling limit and the harmonic oscillator are quite similar. 

Taking Eqs.~(\ref{bas3}) and (\ref{baprox}) into Eq.~(\ref{fop}), one gets, in the weak-coupling limit:
\begin{equation}\label{running}
\hat\Psi^{\hbox{\rm (run)}}(x,t)=\frac{i\delta}{\pi^2}\sqrt{2\over L}\sum_{n\textrm{(odd)}} N^{\hbox{\rm (run)}}(n,x,t)\hat a_n(0)\,,
\end{equation}
where
\begin{equation}\label{fopass}
N^{\hbox{\rm (run)}}(n,x,\tau)\approx\int_{0}^{+\infty}\frac{dy}{n}\frac{e^{-i y^2\tau}\cos(yx/L)}{\left[(y-n\pi)+2i(\delta/2n\pi)^2\right]}\,,
\end{equation}
and $\tau$ is the renormalized time $\tau=\hbar t/2ML^2$.

\begin{figure}
\includegraphics[height=2.4truein,width=3.4truein]{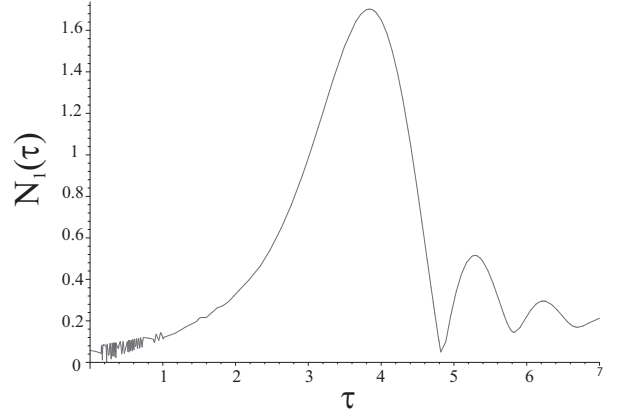}
\caption{\label{fig1}Magnitude $N_1(\tau)=|N^{\hbox{\rm (run)}}(1,x,\tau)|$ of the outgoing wave packet corresponding to the infinite-box level $n=1$, at the position $x=20L$, where $L$ is the box width, as a function of the renormalized time $\tau=\hbar t/2ML^2$.}
\end{figure}

Equations (\ref{running}) and (\ref{fopass}) show that the outgoing field corresponds to a train of wave packets, centered in momentum space around the momenta corresponding to the trap eigenenergies. The expression in Eq.~(\ref{fopass}) is closely related to the Moshinsky function~\cite{Moshinsky,moyses}, which yields the time-dependent behavior of a wave packet initially confined in a half-space: $\psi(x,0)=\theta(-x)\exp(ik x)$, $\Im m\,k<0$, where $\theta(x)$ is the Heaviside function [$\theta(x)=0$ for $x<0$, $\theta(x)=1$ for $x>0$]:
\begin{equation}
M(x,k,t)=\frac{i}{2\pi}\int_{-\infty}^{+\infty}d\kappa\frac{e^{i\kappa x-\hbar \kappa^2t/2m}}{\kappa-k}\,.
\end{equation}
This function exhibits an oscillatory behavior as a function of time, the so-called ``diffraction in time'' effect~\cite{Moshinsky,moyses}. This effect, which has been experimentally observed \cite{Dalibard}, appears when a shutter placed at $x=0$ is opened, letting the initial wave packet, confined to the $x<0$ region, evolve. Here however the integration is from $0$ to infinity, the difference stemming from the fact that in our case the outgoing field emerges from $x=0$ and propagates in both directions. In spite of this difference, we also get here transient effects that can be described in terms of a diffraction in time. This is clearly shown in Fig.~\ref{fig1}, which displays the plot, as a function of the renormalized time $\tau$, of the magnitude of the expression in Eq.~(\ref{fopass}) for $n=1$, $x=20L$, and $\delta^2/2\pi^2=0.2$. In this figure, the small-amplitude fringes close to the origin are due to the interference between the wave packet propagating towards the positive direction with the tail of the packet that propagates in the negative direction. This tail is present even at $t=0$, contrary to what happens with the Moshinsky function, which vanishes exactly for $x>0$ at the initial time. Here, when $t=0$, the sum of the tails of all the wave packets exactly cancels out the bound-state contribution, leading to the vanishing of the source contribution at the initial time.

\subsection{Correlation functions}

From Eqs.~(\ref{operator}) and (\ref{Nfunc}), we can calculate the normal-ordered correlation functions of any order for the system in question. Other orderings could also be considered, by keeping the terms dependent on $\hat b_{k}(0)$ in the above expressions. 

\begin{figure}
\vspace{0.1truein}
\includegraphics[height=4.0truein,width=3.3truein]{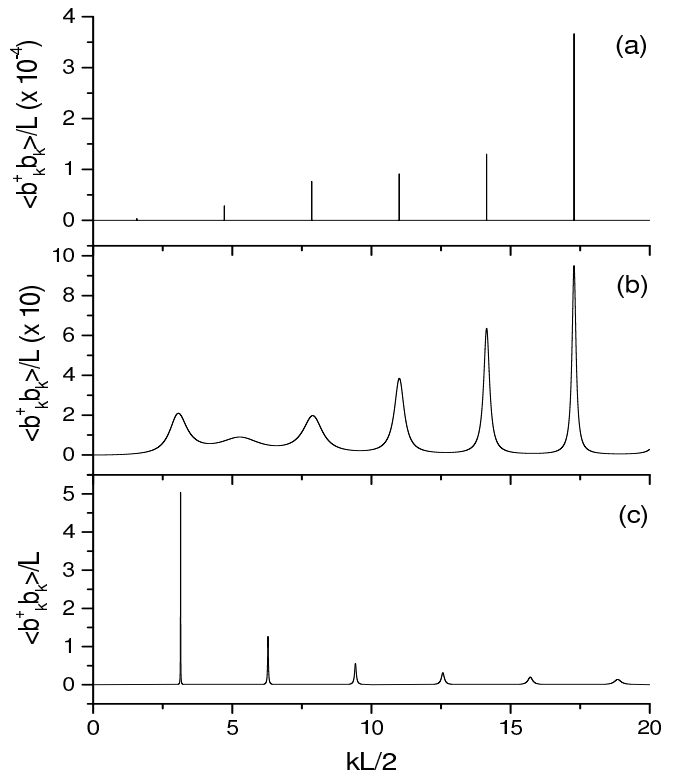}
\caption{Normalized spectral distribution $\langle b^\dagger_k b_k\rangle/L$ of the outgoing fermionic beam, for an infinite-box potential of length $L$, for different coupling strengths: (a) $\delta=0.1$; (b) $\delta=10$; (c) $\delta=100$. Initially, there are 21 atoms in the trap, corresponding to 11 coupled levels. The peaks in (a) are centered around the wave numbers corresponding to the energy levels inside the trap. As the coupling strength  increases [Fig.~(b)], the peaks are displaced towards the situation displayed in Fig.~(c), where the resonances are related to ``dressed states'' of the system trap plus environment. While for weak coupling the height of the peaks increases with the energy, the opposite happens in the strong-coupling limit, when the lowest-energy peak is higher than the others. In these pictures only positive wave numbers are shown, since the complete graphic is symmetric with respect to $kL/2=0$. Also, the totality of eleven peaks in the weak-coupling case is not displayed in Fig.~(a).}
\label{fig2}
\end{figure}

\begin{figure}
\vspace{0.1truein}
\includegraphics[height=4truein,width=3.3truein]{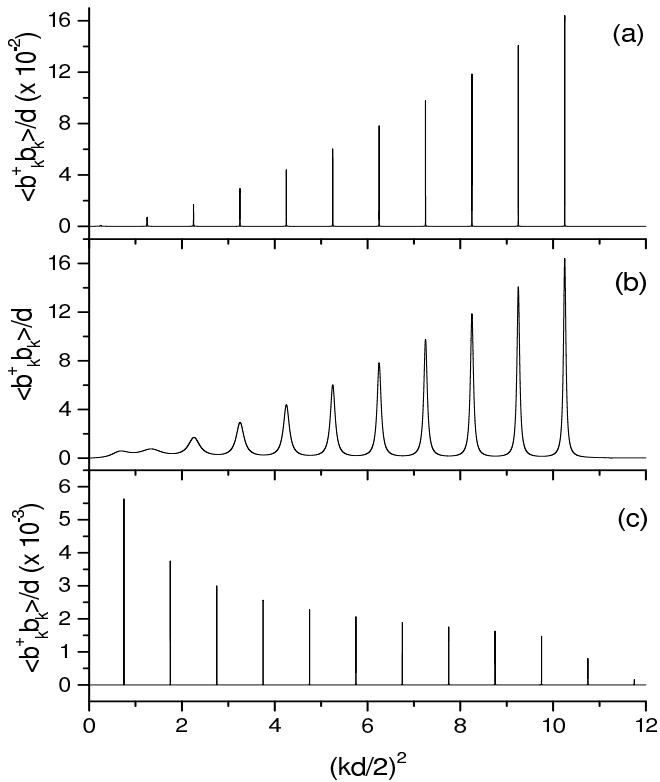}
\caption{Normalized spectral distribution $\langle b^\dagger_k b_k\rangle/d$ of the outgoing fermionic beam, for a harmonic potential, where $d$ is the ground-state width,  for different coupling strengths:  (a) $\delta'=0.1$;  (b)  $\delta'=1$;  (c)  $\delta'=0.1$. The behavior is similar to the one in the previous figure. Fig.~(a) displays eleven peaks, the first one being barely visible and the last one corresponding to the Fermi surface for the $N=21$ trapped atoms. In the strong-coupling regime, one gets instead an infinite number of peaks with decreasing amplitude. The relative importance of the first peak is less pronounced here than in the infinite-box case.}
\label{fig3}
\end{figure} 

Thus, for an initial state diagonal in the number representation, the first-order coherence is given by
\begin{eqnarray}
&&G_{1}(x,x',t)= \langle \hat{\Psi}^\dagger(x,t) \hat{\Psi}(x',t) \rangle \nonumber\\
&&\qquad= \sum_{n} {N}^\ast(n,x,t){N}(n,x',t) \langle \hat a^\dagger_n \hat a_n \rangle (0)\,.
\end{eqnarray}

For $x=x'$, this becomes the beam density $I(x,t)$. The normalized first order correlations functions is defined as
\begin{equation}
g_1(x,x',t)=\frac{G_1(x,x',t)}{\sqrt{I(x,t)I(x',t)}}\,.
\end{equation}

The normal-ordered second-order correlation function,
\begin{equation}
G_2(x,x',t)=\langle \hat{\Psi}^\dagger(x',t) \hat{\Psi}^\dagger(x,t) \hat{\Psi}(x,t) \hat{\Psi}(x',t) \rangle\,,
\end{equation}
may be written, for fermionic atoms, as \cite{baym}:
\begin{equation}
G_2(x,x',t)=I(x,t)I(x',t)-|G_1(x,x',t)|^2\,.
\end{equation}

The normalized second-order correlation function is defined as:
\begin{equation}
g_2(x,x',t)=1-\frac{|G_1(x,x',t)|^2}{I(x,t)I(x',t)}=1-|g_1(x,x',t)|^2\,. \label{2coer}
\end{equation}
The minus sign in the above expression accounts for the anti-bunching property of fermionic beams. This second-order correlation function is analyzed numerically, for the special cases considered in this paper, in Sect.~\ref{numerics}.

For bosonic atoms at zero temperature, the second-order correlation does not depend on the position, while for a thermal distribution it exhibits the bunching effect \cite{baym}.

\begin{figure}
\vspace{0.1truein}
\includegraphics[height=4truein,width=3.3truein]{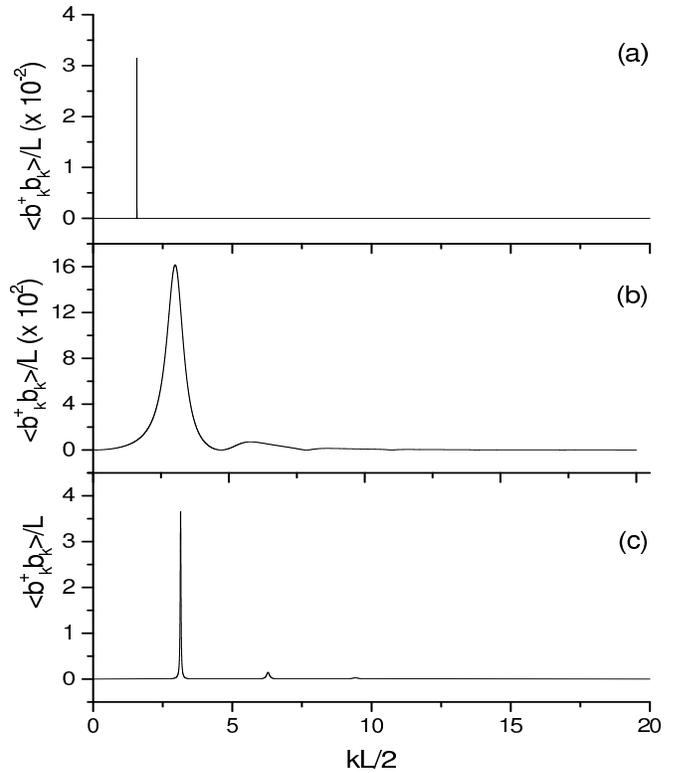}
\caption{Normalized spectral distribution $\langle b^\dagger_k b_k\rangle/L$ of the outgoing bosonic beam, for an infinite-box potential of length $L$, for different coupling strengths: (a) $\delta=0.1$;  (b) $\delta=10$;  (c)  $\delta=100$. For weak coupling -- Fig.~(a) -- there is only one peak, which corresponds to the initially occupied trap level. As the coupling strength increases, new peaks appear, although only the first one remains relatively important. As before, this is an effect of the coupling of the trap levels through the continuum. In this figure only the positive wave numbers are shown, as in Fig.~2.}
\label{fig4}
\end{figure}

\begin{figure}
\vspace{0.1truein}
\includegraphics[height=4.0truein,width=3.3truein]{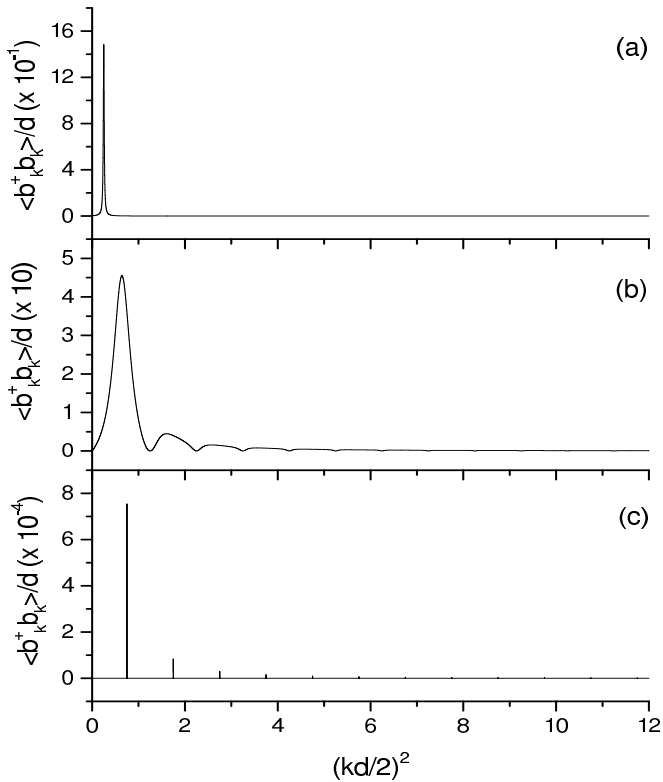}
\caption{Normalized spectral distribution $\langle b^\dagger_k b_k\rangle/d$ of the outgoing bosonic beam, for a harmonic potential, where $d$ is the ground-state width,  for different coupling strengths:  (a) $\delta'=0.1$;  (b)  $\delta'=10$;  (c)  $\delta'=100$. Fig.~(c) clearly displays both the energy displacement and the emergence of new peaks, in the strong-coupling case.}
\label{fig5}
\end{figure}

\begin{figure}
\vspace{0.1truein}
\includegraphics[height=4.0truein,width=3.3truein]{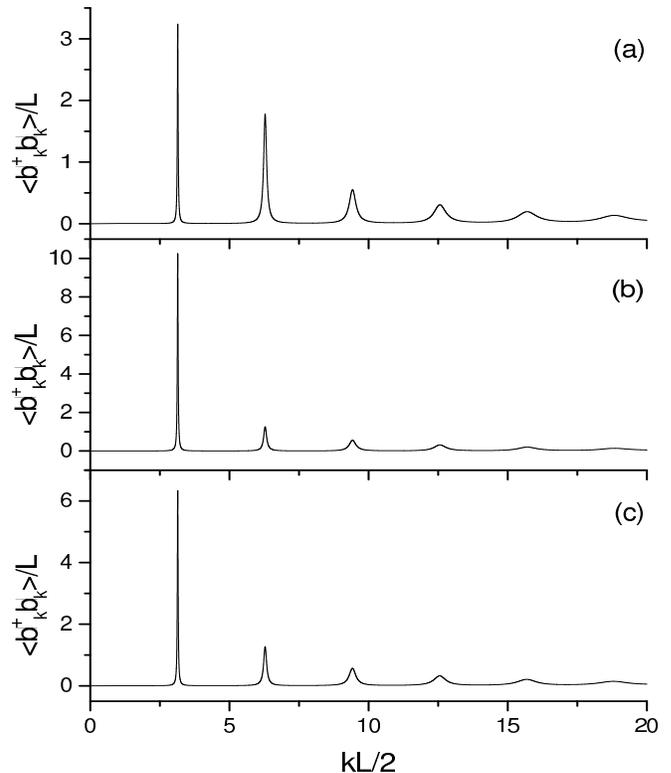}
\caption{Normalized spectral distribution $\langle b^\dagger_k b_k\rangle/L$ of the outgoing fermionic beam, for an infinite-box potential of length $L$, in the strong-coupling regime ($\delta=100$), for the renormalized time $\tau=\hbar t/2ML^2$ equal to: (a) $\tau=0.5$; (b) $\tau=2$; (c) $\tau=5$. For $\tau=10$ one recovers the infinite-time spectral distribution displayed in Fig.~2(c).}
\label{fig6}
\end{figure}

\begin{figure}
\vspace{0.7truein}
\includegraphics[height=3.8truein,width=3.1truein]{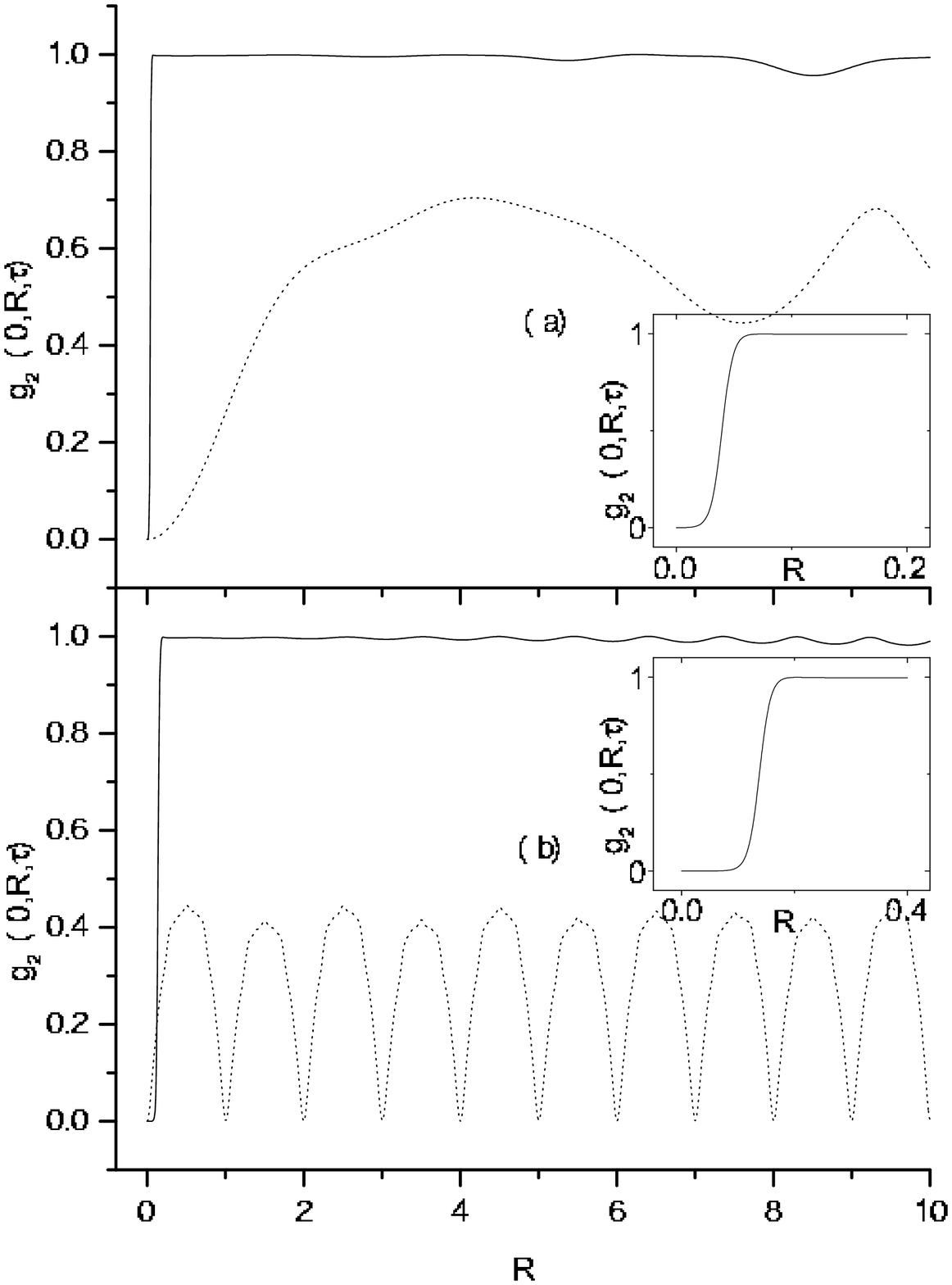}
\caption{Normalized second-order correlation function of the output fermionic beam as a function of the dimensionless position $R$: (a) Harmonic oscillator ($R=x/d$, $d$ being the ground-state width); (b) Infinite-box potential ($R=x/L$, $L$ being the box length). The dimensionless time $\tau$, defined as $\tau=\hbar t/2ML^2$ for the box and as $\tau=2\hbar t/Md^2$ for the harmonic oscillator, is taken equal to 10. The full-line curves (shown in detail in the insets) correspond to $\delta,\delta'=100$, while the dotted curves correspond to $\delta,\delta'=0.1$. Anti-correlation is clearly exhibited in all cases.}
\label{fig7}
\end{figure}

\section{Numerical Results}\label{numerics}

We compute now the behavior of the output atomic beam, for zero temperature. We study the spectral distribution of both a fermionic and a bosonic output beam, and also the second-order correlation function of the outgoing fermionic beam. 

Our main numerical results are shown in Figs.~\ref{fig2} to \ref{fig7}.

\subsection{Spectral distribution of the output beam}

We consider first the spectral distribution $\langle \hat b^\dagger_k(\infty)\hat b_k(\infty) \rangle$ of the output beam in the long-time limit. In the following,  we take the initial number of particles inside the trap to be $N=21$, so that the sums over the contributions of the trap levels will involve eleven terms (since only even states couple to the continuum).  The actual shape of the spectral distribution depends strongly on the value of the adimensional parameters $\delta$ and $\delta'$, defined respectively by Eqs.~(\ref{delta}) and (\ref{delta'}).  The spectral distribution for fermions trapped by an infinite box is displayed in Fig.~\ref{fig2}, while the distribution corresponding to a harmonic potential is displayed in Fig.~\ref{fig3}. 

We can see that for small values of $\delta$ and $\delta'$, the peaks in the spectral distribution can be simply interpreted as resonances associated with the unperturbed trap levels corresponding to even eigenstates, consistently with the previous analysis. One should notice that the height of the peaks grows with $k$ (the widths of the peaks cannot be seen, within the scale of the figures). In fact, in the weak-coupling limit the approximate expressions given by Eqs.~(\ref{weaksp}) and (\ref{weakspoh}) lead to excellent fits to the spectra exhibited in Figs.~\ref{fig2} and \ref{fig3}. 

For larger values of $\delta$ and $\delta'$, the peaks are shifted towards the strong-coupling dressed energies, as discussed in Sections \ref{sbox} and \ref{sho}. Another important feature, also discussed in those Sections, is the relative amplitude of the peaks. We see that, in the strong-coupling limit, the first peak is the highest one,  the amplitude of the successive peaks decreasing now with the energy. This feature is especially pronounced for the infinite box. In this case, we have the remarkable feature that the outgoing beam has an accentuated monochromatic character, in spite of the multitude of bound levels of the infinite box. Again, there is excellent agreement of the strong-coupling plots with the approximate expressions given by Eqs.~(\ref{strsp}) and (\ref{strspoh}). 

In Figs.~\ref{fig4} and \ref{fig5}, we study the behavior of the corresponding bosonic systems, initially in the ground state. For small values of $\delta$ and $\delta'$, the peak is related to the only occupied level of the system. For stronger couplings, new peaks appear, around energy levels corresponding to the excited trapped states, which get populated as a consequence of the interaction between the ground state and the continuum. We note that this effect is absent in models that treat the bosonic system as a single trapped level.

The results shown so far correspond to the spectral distribution of the outgoing atoms in the infinite-time limit. The behavior of the time-dependent spectral distribution for finite times is very simple in the case of weak coupling: then, each level behaves independently of the others, so that as the population of each trap level decays, as discussed in Section \ref{population}, the population of the corresponding free-space mode increases, finally getting to the distributions displayed in Figs.~\ref{fig2} and \ref{fig3}. 

For strong coupling, the dynamics of the spectral distribution is more involved, since now the different trap levels interact with each other through the continuum, and there is a strong probability, for finite times, that the atoms are reabsorbed into the trap (this phenomenon shows up in the oscillatory behavior of the atomic population in the trap). Figure 6 exhibits the approach to the infinite-time limit of the time-dependent spectral distribution, for the special case of strong coupling, and for an infinite-box potential. One should note that the relative heights of the spectral peaks are time-dependent, their precise relationship being related to the intricate transient behavior of the system.

\subsection{Second-order correlation for fermions}

Figure \ref{fig7} displays the behavior of the second-order correlation function as a function of the distance, for the fermionic case . The anti-correlation of fermions is clearly exhibited in all cases. For strong coupling, this correlation function goes very fast to one. 

\section{Conclusions}

We have shown that the dynamics of trapped fermions with an output coupling may exhibit very interesting features. The main source of these features, which is also the most challenging aspect of this problem, as compared to the corresponding situation for bosons at zero temperature, is the multitude of energy levels of the trap that are necessarily populated, due to the Pauli principle. 

In view of the complexity of the problem, our strategy in this paper was to deal with a model simple enough so that it was possible to obtain some analytic handle on it, yet sufficiently rich to demonstrate interesting features of this system. The assumption of a delta-type coupling indeed greatly simplified the solution of the problem, while still keeping the main feature of leading to a coupling between the trap eigenstates mediated by the continuum. This coupling has remarkable effects in the strong-coupling limit, leading to clear signatures of dressed energies in the infinite-time spectral distribution of the output beam. The same kind of coupling is present in the bosonic case, albeit its effects are less dramatic if only the ground state is initially populated (which is the situation when the temperature is zero). 

A peculiar characteristic of the system here considered, also present in one-dimensional single-mode boson models, is the presence of a bound mode of the coupled system, for any value of the coupling constant. This implies a non-Markovian behavior of this system, and has two important consequences: the probability of finding the atoms in the trap is oscillatory, and a fraction of the atoms remains in the trap, even in the infinite-time limit. For strong coupling, this fraction approaches $1/4$ when the number of atoms is much larger than one. This bound mode should be however highly sensitive to an external potential like a gravitational field. 

Since in the model here considered there is no external replenishing of the trap, the outgoing beam has a non-stationary nature. For finite times, it displays a very intricate dynamics, which results from the combined effect of a train of wave packets, with transient behavior that exhibits the so-called "diffraction in time" effect, and which overlap with the bound-state wave function.

In the infinite-time limit, however, it is possible get an analytical expression for the spectral distribution of the outgoing atoms. In the strong-coupling limit, and for a steep trapping potential, the outgoing atomic beam exhibits remarkable features, for large times: it is quasi-monochromatic, and it displays anti-bunching. It is interesting to remark that the combination of these two features is highly desirable, although hard to achieve, in light beams. Indeed, the generation of low-noise laser light has been an intense field of research~\cite{Yamamoto}, since the first experimental observations of anti-bunching \cite{Kimble} and sub-Poissonian statistics \cite{Mandel}. For the fermionic beams considered here, anti-bunching comes out quite naturally. On the other hand, we have shown that, under certain conditions, it is also possible to get here, in the infinite-time limit, a quasi-monochromatic spectral distribution, in spite of the large number of occupied energy levels in the trap. This could be especially helpful for some applications recently envisaged for fermionic atomic beams, like for instance the development of low-noise atomic interferometers~\cite{Meystre2}.

The investigation of more realistic situations, including for instance the presence of a gravitational field, will be the object of further consideration.

\begin{acknowledgments}
This work was partially supported by PRONEX (Programa de Apoio a
N\'ucleos de Excel\^encia), CNPq (Conselho Nacional de Desenvolvimento
Cient\'\i fico e Tecnol\'ogico), FAPERJ (Funda\c c\~ao de Amparo \`a Pesquisa do Estado do Rio de Janeiro), FUJB (Funda\c c\~ao Universit\'aria
Jos\'e Bonif\'acio), and the Millennium Institute on Quantum Information.
\end{acknowledgments}

\appendix

\section{Sum for the Infinite Box}

\label{ap1}

In this appendix, we calculate the sum in Eq.~(\ref{fmu}). 

From Ref.~\cite{knopp} we have the following result:
\begin{equation} 
\pi\cot{(\pi z)}=\frac{1}{z}+\sum_{m=1}^{\infty}\Big[ \frac{1}{z-m} + \frac{1}{z+m} \Big]\,,
\end{equation}
so that 
\begin{equation}
\pi\cot{(\pi z)}=z\sum_{m=-\infty}^{\infty} \frac{1}{z^2-m^2}\,.  \label{sumcot}
\end{equation}

In our case, we have the sum:
\begin{equation}
S=\sum_{m=1 \hbox{\rm(odd)}}^{\infty} \frac{1}{z^2-m^2}\,,
\end{equation}
which may be written as
\begin{equation}
S=\sum_{m=1}^{\infty} \frac{1}{z^2-m^2}-\sum_{m=1 \textrm{(even)}}^{\infty} \frac{1}{z^2-m^2}\,.
\end{equation}

Setting in the second sum $m=2n$, we may write:
\begin{equation}
S=\frac{1}{2}\Big[\sum_{m=-\infty}^{\infty} \frac{1}{z^2-m^2}-\sum_{n=-\infty}^{\infty} \frac{1}{z^2-(2n)^2}\Big]\,,
\end{equation}
so that, from Eq.~(\ref{sumcot}),
\begin{equation}
S=\frac{\pi}{2}\Big[\frac{\cot{(\pi z)}}{z}-\frac{\cot{(\pi z/2)}}{2z} \Big]\,.
\end{equation}

Since
\begin{equation}
\cot{(x)}=\frac{1}{2}\left[ \cot{\left( \frac{x}{2}\right)}-\tan{\left( \frac{x}{2}\right)}   \right]\,,
\end{equation}
we finally obtain:
\begin{equation}\label{finalsum}
\sum_{m=1\,\textrm{(odd)}}^{\infty} \frac{1}{z^2-m^2}=-\frac{\pi}{4}\frac{\tan(\pi z/2)}{z}\,,
\end{equation}
which leads to Eq.~(\ref{fomega}).

Also, letting $z\rightarrow iz$, we get:
\begin{equation}\label{finalsum2}
\sum_{m=1\,\textrm{(odd)}}^{\infty} \frac{1}{z^2+m^2}=\frac{\pi}{4}\frac{\tanh(\pi z/2)}{z}\,,
\end{equation}
which when applied to Eq.~(\ref{fmu}) leads to Eq.~(\ref{resfmubox}).

\section{Sum for the Harmonic Oscillator}
\label{ap3}

In this Appendix, we evaluate the sum in Eq.~(\ref{resfmuho}). We start by proving the identity
\begin{equation}
S=\sum_{m=0}^{\infty} \frac{|\varphi_{2m}(0)|^2}{z+m}=\frac{1}{\sqrt{\pi d^2}}\times \frac{\sqrt{\pi}\Gamma(z)}{\Gamma(z+1/2)}\,, \label{sumap}
\end{equation}
where $\Gamma(z)$ is the Gamma function, $d=\sqrt{\hbar/m\omega_0}$ is the width of the ground state of the harmonic oscillator, and, from Eq.~(\ref{wavefuncap}), 
\begin{equation}
\varphi_{2m}(0)=\big(\frac{1}{\pi d^2} \big)^{1/4} \frac{1}{\sqrt{2^{2m} (2m)!}} H_{2m}(0)\,,
\end{equation}
where $H_n(x)$ is the Hermite polynomial of order n, with (see ref. \cite{abramowitz}, p. 777)
\begin{equation}
H_{2m}(0)=(-1)^m \frac{(2m)!}{m!}\,.
\end{equation}

Replacing these two last expressions into Eq.~(\ref{sumap}), we obtain
\begin{equation}
S=\frac{1}{\sqrt{\pi d^2}} \sum_{m=0}^{\infty} \frac{(2m)!}{2^{2m}(m!)^2}\frac{1}{z+m}\,.
\end{equation}

Let us consider now the function $M(z)=\sqrt{\pi}\Gamma(z)/\Gamma(z+1/2)$, and prove that $S=M(z)$. The function $M(z)$ is a meromorphic function, with poles on the non-positive integers in the complex plane: $z=-m$, $m=0,1,2,3,\dots$. Therefore, we may write \cite{copson}:
\begin{equation}
M(z)=\sum_{m=0}^{\infty} \frac{[\textrm{Residue of M(z) for}\, z=-m] }{z+m}\,. \label{sum2}
\end{equation}

The residues in the above equation are given by:
\begin{eqnarray}
\textrm{Res}[M(z),-m]&=&\lim_{z\rightarrow-m}(z+m) \frac{\sqrt{\pi}\Gamma(z)}{\Gamma(z+1/2)}\nonumber \\ &=&\frac{\sqrt{\pi}(-1)^m}{m!\Gamma(-m+1/2)}\,. \label{resg}
\end{eqnarray}

We use now that:
\begin{equation}
\frac{1}{\Gamma(1/2-m)}=\frac{\sin{(\frac{\pi}{2}-m\pi)}\Gamma(m+1/2)}{\pi}\,, \label{gmmeio}
\end{equation}
\begin{equation}
\Gamma(m+1/2)=\frac{(2m)! \Gamma(1/2)}{2^{2m} m!}\,, \label{gmeio}
\end{equation}
and  $\Gamma(1/2)=\sqrt{\pi}$, and replace Eqs.~(\ref{resg}), (\ref{gmmeio}), and (\ref{gmeio}) into Eq.~(\ref{sum2}), obtaining finally:
\begin{equation}
\frac{\sqrt{\pi}\Gamma(z)}{\Gamma(z+1/2)}=\sum_{m=0}^{\infty} \frac{(2m)!}{2^{2m}(m!)^2}\frac{1}{z+m}\,,
\end{equation}
which proves the desired identity.

It follows then immediately that
\begin{equation}
 \sum_{m=0}^{\infty} \frac{|\varphi_{2m}(0)|^2}{m+1/4 +(\mu^2/2\omega_0) }=\frac{1}{d}\frac{\Gamma(1/4+\mu^2/2\omega_0)}{\Gamma(3/4+\mu^2/2\omega_0)}\,.
\end{equation}

\end{document}